# All-dielectric periodic terajet waveguide using an array of coupled cuboids

I.V. Minin,[1,a)] O. V. Minin, [1,b)] V. Pacheco-Peña, [2,c)] M. Beruete [2,d)]

[1]Siberian State Institute of Metrology, Dimitrove 4, Novosibirsk, 630004, Russia
[2]TERALAB (MmW—THz—IR &PlasmonicsLaboratory), Universidad Pública de Navarra, Campus Arrosadía, 31006 Pamplona, Spain

In this paper, the recently proposed technique to produce photonic jets (terajets at THz frequencies) using 3D dielectric cuboids is applied in the design of mesoscale cuboid-chain waveguide. The chains are basically designed with several dielectric cubes with dimensions $\lambda_0$ along the *x*, *y* and *z* axes placed periodically along the axial z-axis and separated by an air-gap. Based on this, a systematic study of the focusing properties and wave guiding of this chain is performed when the air-gap between the dielectric cubes is changed from $0.25\lambda_0$ to $2.5\lambda_0$ with the best performance achieved with the latter design. The numerical results of focusing and transport properties are carried out using Finite Integration Technique. The results here presented may be scaled to any frequency ranges such as millimeter, sub-millimeter or optical frequencies.

[a)] Electronic mail: prof.minin@gmail.com
[b)] Electronic mail: prof.minin@gmail.com
[c)] Electronic mail: victor.pacheco@u navarra.es
[d)] Electronic mail: miguel.beruete@unavarra.es

## I. Introduction

Probably the first experimental prototype of a lens beam waveguide with a set of identical lenses was designed by Christian and Goubau in 1961 for a wavelength of 1.25 cm and clearly demonstrated its potential[1]. In this system, several dielectric lenses with diameter of 20 cm made of expanded polystyrene were used periodically spaced by 1 m. To reduce absorption losses in lens materials the diffractive lens[2] was first used in quasi-optical transmission lines by Sobel *et al.* in 1961[3]. It was demonstrated that the transmission line had low losses at 210 GHz (only 2 dB over a distance of 17 m).

As it is well known, common lens transmission lines are designed by using lenses with thickness much smaller that its diameter. Based on this, several types of lens transmission lines have been reported using dielectric lenses[4]. On the other hand, at optical frequencies, it has been also demonstrated that linear arrays of dielectric spheres can operate as lossless waveguides for beams with certain spatial properties which depend on the relative size of the dielectric spheres.[5]

The periodically focused modes (PFM) concept has been stimulated by studies of the chains designed with mesoscale spheres (i.e., diameters of particles of $4\lambda_0 < D < 10\lambda_0$, where $\lambda_0$ is the working wavelength)[6] and cylinders[7]. These studies have been performed in the context of coupling between their whispering gallery modes due to the beam focusing produced by individual spheres and micro-disks when they are illuminated with a planewave; i.e., photonic jets or nanojets.[8-11] This can be achieved due to the fact that the Brewster angle conditions are periodically reproduced for PFMs with a period matching the size of two sphere diameters, and the coupling between nanojets in chains of spheres[12]. The latter mechanism leads to formation of the so-called periodic nanojet-induced modes observed in chains of polystyrene microspheres[13-17], cylinders and clusters[18] of polystyrene microspheres, in which the nanojets can be periodically reproduced along a chain of microspheres. These modes result from the optical coupling of microspheres acting as a series of micro-lenses, which periodically focus propagating waves into photonic nanojets. It has been demonstrated that the coupled focused beams decrease in size along the chain of polystyrene microspheres with index $n$ =1.59, reaching wavelength scale dimensions in the case of small beads with $4 < D/\lambda_0 < 10$.[19] It must be noted that all dielectric spheres in such linear arrays are located close to each other and the elementary "unit-cell" of the array consists of two spheres.

Cubic dielectric particles are promising candidates for low-loss, ultra-compact, photonic applications owing to their relative ease of fabrication as compared to spherical particles. By using these structures, the capability to produce jets at terahertz (THz) and sub-THz frequencies (so-called terajets, analog of photonic jet in optics) has been recently proposed and demonstrated experimentally by using 2D and 3D dielectric cuboids.[20,21]

In this communication, the capability to produce periodically induced terajets along a chain of mesoscale 3D dielectric cubes (here, at difference with our previous works we use regular hexahedrons) separated with air-gaps of different length is numerically studied. Numerical results demonstrate that the optimum air-gap length is $2.5\lambda_0$. Moreover, the influence of dielectric losses in the wave guiding performance is evaluated by introducing overestimated values of loss tangent ($\delta$), demonstrating that the structure here proposed has a robust performance, due to the fact that the spatial resolution of the terajets produced at the output surface of each cube of the chain is not strongly deteriorated when losses are increased. The results here presented may be directly scaled to other frequency bands such as optical frequencies. All the numerical results were performed using the Finite Integration Technique with the transient solver of the commercial software CST Microwave Studio™ with the same boundary conditions and mesh as in[20–22].

The 3D elementary dielectric cube used in this letter is schematically shown in Fig. 1(a). It is a regular hexahedron of side $L = \lambda_0$, with refractive index $n = 1.46$ and immersed in vacuum ($n_0 = 1$). In order to design a chain, several cubes (three in this case) are placed periodically along the optical z-axis separated by an air-gap, as shown in Fig. 1(b). The whole structure is illuminated from the left side by using a vertically polarized plane wave ($E_y$) with its propagation direction along the optical z-axis.

Based on this, we first evaluate the focusing performance and wave guiding properties of the chain when the distance between cubes is changed. Numerical results of the normalized power distribution on the E-plane (yz-plane) for values of the air-gaps between cubes varying from $0.25\lambda_0$ to $2.5\lambda_0$ are shown in Fig. 2. When the air-gaps are smaller than $\lambda_0$ the terajets are produced inside the cubes [Fig. 2 (a-c)] and the higher power distribution is obtained at the central cube (close to its output surface). In these cases, the terajets are somewhat blurred for the first and third cubes. On the other hand, when the air-gaps increase terajets are clearly formed just at the output surface of each cube. Note that the best performance is achieved when an air-gap of $2.5\lambda_0$ is used, where the power distribution of the terajets is increased for all the cubes [See Fig. 2(f)]. In order to better compare the performance, simulation results of the normalized power distribution along the optical z-axis air-gap values equal to $0.25\lambda_0$ and $2.5\lambda_0$ [taken from Fig. 2 (a) and (f)], are shown in Fig. 3. Here it is clearly observed that the best performance is achieved when the air-gap is $2.5\lambda_0$ where the maxima is obtained just at the output surface of each cube, demonstrating that the terajets are efficiently excited with this configuration. It is important to highlight that several simulations with air-gaps up to $4.5\lambda_0$ were performed and the terajets were also observed just at the output of each cube (not shown here). As it has been pointed out previously, the power distribution has a maximum value at the central cube compared with the first one. This is due to the fact that reflection losses are partially compensated by the flat surface of the cubes at the back and front sides. Also, note that the air-gap has an important role in the performance of the chain. This is due to its value should be, at least, the same as the length of the terajet in order to couple one terajet to the next one. This is indicative of the formation of terajet-induced modes having the smallest propagation losses in chains of cubes. This performance is degraded for the smaller gaps since the terajets are not completely formed in such small air-gaps and, therefore, it is not able to couple to the next cube.

For the sake of completeness, numerical results of the full-width at half-maximum (FWHM) along both transversal *x*- and *y*- axes at each terajet for the case with air gap $2.55\lambda_0$ are shown in Fig. 4. Here the cube number 1, 2 and 3 corresponds to the left, central, and right cube of the chain. It can be observed that a similar resolution is achieved for all the terajets along both directions with values below $0.55\lambda_0$ for all cases. Based on this, the ellipticity (defined as the ratio between both transversal resolutions $FWHM_x/FWHM_y$)[22] of all terajets are close to 1. Therefore, a quasi-spherical spot is obtained for the chain configuration in good agreement with our previous results dealing with isolated cuboids [20–22]. Regarding the exploration range ($\Delta z$) [20,21], also shown in Fig. 4, it can be observed that the maximum value is obtained for the first cube while it is reduced for the rest. For this chain, the power enhancement (defined as the power at the output surface of each cube compared with the power without cubes) is ~8.5, ~11.6 and ~10.8 (in linear scale) for the first, second and third cube, respectively, in good agreement with the results of a single cuboid [20]. Thus, it is observed that a mesoscale dielectric cube-air-chain waveguide is mainly characterized by the periodicity of jets corresponding to the optimal length of air-gap between cubes when a fixed geometry and materials (refractive index contrast between the background medium and the cubes[20]) of the cubes is chosen.

This fact contrasts with the results found in optical chains of dielectric microspheres where the periodical focusing of light in straight chains of touching microspheres is characterized with the periodicity of photonic nanojets corresponding to the size of two spheres.[15]

Finally, it is important to evaluate the performance of the cube-air chain by taking into account the dielectric losses due to the absorption of dielectric materials at THz frequencies. Numerical results of the normalized power distribution on the $E(yz)$-plane when periodicity is $2.5\lambda_0$ and the cubes have a loss tangent of $\delta = 0.05$ and $\delta = 0.1$ at the working wavelength $\lambda_0$ are shown in Fig. 5(a) and (b), respectively. Note that these values are overestimated since, for example, a typical value $\delta$ for Teflon (with a similar refractive index as the one here used) is ~0.008 at 1 THz.[23] To better compare these results, the power distribution along the optical $z$-axis at $x = 0$ is shown at the bottom of each plot for both cases. It can be observed that the terajets are slightly shifted away from the output surface of each cube when losses are introduced. Moreover, the power distribution at each output is reduced up to ~1/3 when $\delta = 0.1$ (compared with the value obtained when no losses are included). These results are not surprising since a change in $\delta$ modifies the refractive index of the cubes and therefore a shift of the focus should occur[20]. Also the length of terajet is increased when increasing the loss tangent. To better compare these results the focusing properties of the terajets produced at the output surface of each cube for these values of $\delta$ are summarized in Table 1. In general, it is shown that the resolution of each terajet is deteriorated when losses are increased. However, the results obtained are very close to the lossless cases even when the dielectric losses here evaluated are overestimated. For the sake of completeness, the normalized power distribution along the z-axis for two chains of 11 cuboids with the previous values of $\delta$ is shown in Fig. 5(c). It can be observed that the power decays exponentially at half its maximum at the fifth cuboid. Moreover, it is shown that, from the fifth cuboid, the relative power at each terajet still decays but with small losses (~0.26dB/cuboid and ~0.4dB/cuboid for the cases with $\delta = 0.1$ and $\delta = 0.05$, respectively), compared with the first five terajets, and the terajets are excited in the rest of the cuboids which form the chain for both values of $\delta$. It is important to note that these values have been calculated from the normalized power distribution shown in Fig. 5(c), which is normalized to the maximum from each case. Therefore, even when the decay of power at each terajet position is lower for the case with $\delta = 0.1$, the total power at the terajet maxima are lower than those obtained with $\delta = 0.05$, as it has been explained before and can be corroborated in the bottom panel of Fig. 5(a) and (b). The ratio of the power maximum for the cases with $\delta = 0.05$ and $\delta = 0.1$ is ~1.9 dB.

Finally, regarding the FWHM along the transversal *x*-axis for the first, fifth and last cuboid of Fig. 5(c), the resulting values are $0.49\lambda_0$, $0.48\lambda_0$ and $0.49\lambda_0$, respectively, for the case with $\delta = 0.05$ while they are $0.5\lambda_0$, $0.49\lambda_0$ and $0.5\lambda_0$, respectively, for the case with $\delta = 0.1$. Demonstrating that the spatial resolution of the terajets is maintained along both chains. Based on these results, the robustness of the wave guiding configuration here proposed when increasing losses is therefore demonstrated.

TABLE I. Numerical results of the performance for each terajet produced in a cube-air-chain with an air gap of $2.5\lambda_0$ for different values of loss tangent $\delta$.

| Cube number | FWHM$_x$($\lambda_0$)[a] | | FWHM$_y$($\lambda_0$)[b] | | $\Delta_z$($\lambda_0$)[c] | |
|---|---|---|---|---|---|---|
| | $\delta=0.05$ | $\delta=0.1$ | $\delta=0.05$ | $\delta=0.1$ | $\delta=0.05$ | $\delta=0.1$ |
| 1 | 0.49 | 0.5 | 0.45 | 0.47 | 1.1 | 1.37 |
| 2 | 0.5 | 0.51 | 0.43 | 0.45 | 0.97 | 1.71 |
| 3 | 0.49 | 0.5 | 0.43 | 0.44 | 0.83 | 1.01 |

[a]FWHM$_x$ is the Full-width at half-maximum along the *x*-axis.
[b]FWHM$_y$ is the Full-width at half-maximum along the *y*-axis
[c]$\Delta_z$ is the photonic jet exploration range.

In conclusion, the capability to design cube-air-chains waveguides using 3D dielectric cubes of refractive index $n = 1.46$ immersed in vacuum ($n_0 = 1$) has been studied. Basically, the waveguide has been designed with a periodic series of 3D dielectric cubes with an air-gap between the cubes. A systematic study of the wave guiding performance has been shown by changing the air-gap between the cubes from $0.25\lambda_0$ up to $2.5\lambda_0$ using three cubes along the optical *z*-axis. This wave guiding stems from the quasi-optical coupling of mesoscale cubes acting as a series of flat lenses, which periodically are able to focus the propagating waves generating terajets at their output faces. Based on this, it has been shown that the terajets are not produced for air-gaps $< \lambda_0$ therefore no wave guiding is achievable using these configurations. However, by increasing the air-gap, the terajets are generated at the output surface of each cube with an optimum value of air-gap of $2.5\lambda_0$. The focusing properties of the terajets produced at the output surface of each cube of the system has been studied in terms of FWHM along both transversal *x*- and *y*- axes, terajet exploration range and enhancement. Numerical results demonstrate that values of FWHM below $0.5\lambda_0$ are achievable for all the cubes along both transversal directions. Therefore a quasi-spherical terajet is produced with this configuration due to the ellipticity is very small (close to 1). Also, it has been demonstrated that the power enhancement achieved with this design is maximum for the terajet produced at the central cube of the chain with a value of ~11.6.

Finally the influence of dielectric losses has been evaluated in terms of both transversal resolution and exploration range by introducing two overestimated values of loss tangent (δ = 0.05 and δ = 0.1). It has been shown that the terajets are moved away from the output surface of each cube as expected and the power at each focal position is reduced up to ~1/3 compared to the value received when no losses are included. However, even when higher dielectric losses are introduced in the system, the resolution of the terajets is not strongly deteriorated demonstrating the robustness of the wave guiding configuration here proposed. Since all the studies here presented have been performed using normalized dimensions for the whole structures, the wave guiding using periodic arranged mesoscale cubes with air gaps may be scaled to other frequency bands such as millimeter, sub-millimeter and optical frequencies and may be applied to microscopy applications and devices such as power dividers and multiplexers.

**ACKNOWLEDGMENTS**

This work was supported in part by the Spanish Government under contract TEC2011-28664-C02-01. V.P.-P.is sponsored by Spanish Ministerio de Educación, Cultura y Deporte under grant FPU AP-2012-3796. M.B. is sponsored by the Spanish Government via RYC-2011-08221.

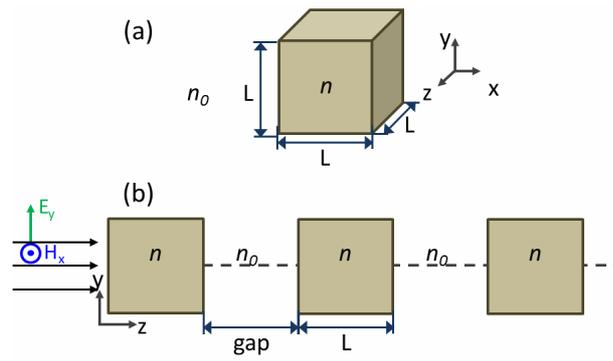

FIG. 1. (a) Schematic representation of a 3D dielectric cube with side $L = \lambda_0$. The 3D cube has a refractive index $n = 1.46$ and is immersed in vacuum ($n_o=1$). (b) Schematic representation of a chain of 3D dielectric cubes immersed in vacuum with an air-gap between them.

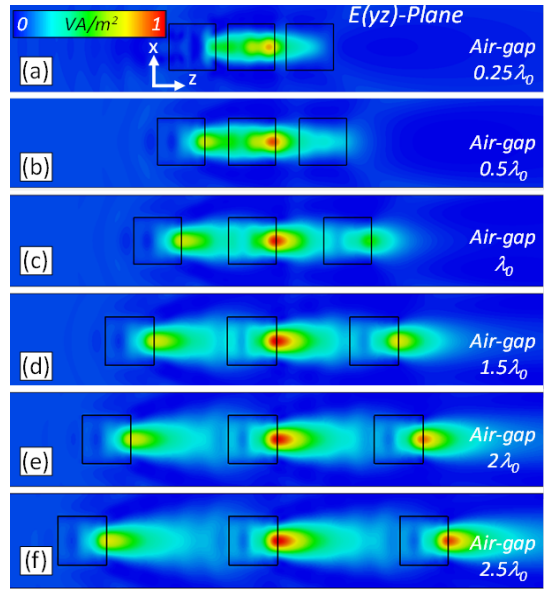

FIG. 2. Numerical simulations of the power distribution on the *yz*-plane/*E*-plane for several chains of 3D dielectric cuboids with dimensions $L = \lambda_0$ along all directions when the air-gap between each cuboid is selected as: (a) $0.25\lambda_0$, (b) $0.5\lambda_0$, (c) $\lambda_0$, (d) $1.5\lambda_0$, (e) $2\lambda_0$ and (f) $2.5\lambda_0$. The numerical results have been normalized with respect to the maximum power distribution achieved from all the structures which in this case corresponds to the design with an air-gap of $2.5\lambda_0$.

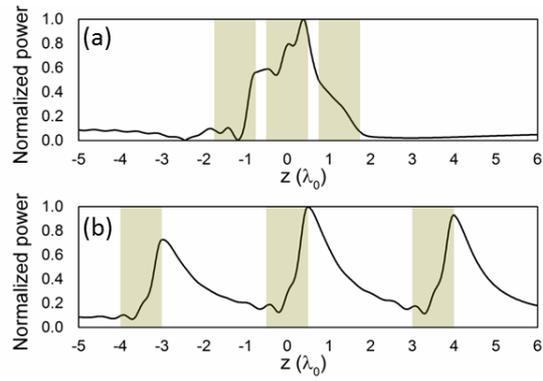

FIG. 3. Simulation results of the normalized power distribution along the optical $z$-axis the two chains of 3D dielectric cuboids shown in Fig. 2, when the air-gap between each cuboid is selected as (a) $0.25\lambda_0$ and (b) $2.5\lambda_0$. Both results are normalized with respect to the maximum power of each case.

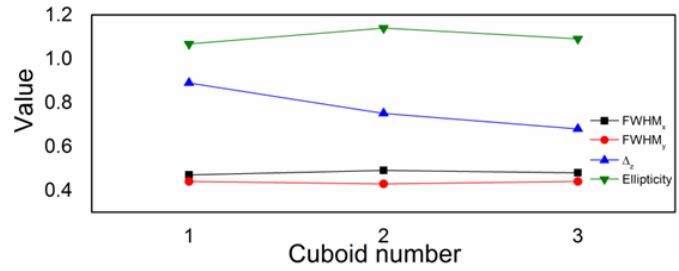

FIG. 4. Numerical results (in wavelengths λ₀) of the Full-width at half maximum along the *x*- (black curve ) and *y*-axes (red curve) at each output surface of each cuboid. Exploration range in wavelengths ($\Delta_z$) for each terajet produced at the output surface of each cuboid (blue curve) along with the ellipticity for each terajet (red curve). All curves have been connected with a continuous line to guide the eye.

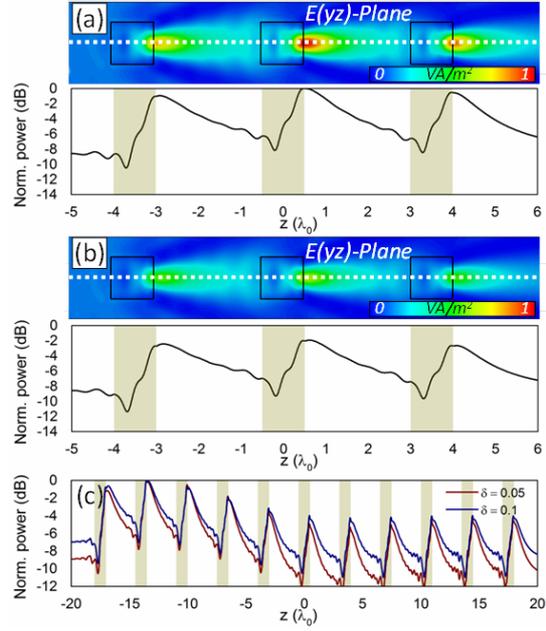

FIG. 5 Numerical simulations of the power distribution on the *E(yz)*-plane a chain of 3D dielectric cuboids with dimensions *L*= $\lambda_0$ along all directions with an air-gap between each cuboid with a value of 2.5$\lambda_0$ when dielectric losses are introduced to the cuboids with a value of: (a) δ=0.05 and (b) δ=0.1. The bottom plot of each panel represents the power distribution along the optical *z*-axis along the dashed white line for each case. The normalized values are calculated using the maximum from all cases, which corresponds to the terajet produced at the output surface of the second (central) cuboid with δ=0.05. (c) Normalized power distribution along the optical *z*- axis for a chain of 11 dielectric cuboids with the same dimensions as (a) and (b) for the two values of loss tangent δ=0.05 (red line) and δ=0.1 (blue line). The normalized values are calculated based on the maximum of each case.